\documentclass[twoside]{sfm}

\input{sf.def}
\usepackage{natbib}

\begin{document}

\newcommand{\lxlb}{$L_{\rm X}/L_{\rm BOL}$}
\newcommand{\loglxlb}{$\log[L_{\rm X}/L_{\rm BOL}]$}
\newcommand{\cz}{\ensuremath{C_Z}}
\newcommand{\pv}{\ensuremath{P_V}}
\newcommand{\nv}{\ensuremath{N_V}}
\newcommand{\pvr}{\ensuremath{P_V^\mathrm{(rect)}}}
\newcommand{\nvr}{\ensuremath{N_V^\mathrm{(rect)}}}
\newcommand{\bz}{\ensuremath{\langle B_z\rangle}}
\newcommand{\sbz}{\ensuremath{\sigma_{\langle B_z\rangle}}}
\newcommand{\nz}{\ensuremath{\langle N_z\rangle}}
\newcommand{\snz}{\ensuremath{\sigma_{\langle N_z\rangle}}}
\newcommand{\fo}{\ensuremath{f^\parallel}}
\newcommand{\fe}{\ensuremath{f^\perp}}
\def\llm{{\sc LLmodels}}
\def\atl{{\sc ATLAS9}}
\def\aatl{{\sc ATLAS12}}
\def\starsp{{\sc STARSP}}
\def\aur{$\Theta$~Aur}
\def\logg{\log g}
\def\tauros{\tau_{\rm Ross}}
\def\kms{km\,s$^{-1}$}
\def\bz{$\langle B_{\rm z} \rangle$}
\def\degr{^\circ}
\def\aaps{A\&AS}
\def\aap{A\&A}
\def\apjs{ApJS}
\def\apj{ApJ}
\def\rmxaa{Rev. Mexicana Astron. Astrofis.}
\def\mnras{MNRAS}
\def\actaa{Acta Astron.}
\newcommand{\Tef}{T$_{\rm eff}$~}
\newcommand{\Vt}{$V_t$}
\newcommand{\CC}{$^{12}$C/$^{13}$C~}
\newcommand{\CDC}{$^{12}$C/$^{13}$C~}

\pagebreak

\thispagestyle{titlehead}

\setcounter{section}{0}
\setcounter{figure}{0}
\setcounter{table}{0}

\markboth{Naz\'e}{Magnetic fields in O stars}

\titl{Magnetic fields in O stars}{Ya\"el Naz\'e$^1$}
{$^1$FNRS-Dept AGO, ULg, Li\`ege, Belgium, email: {\tt naze@astro.ulg.ac.be} }

\abstre{Over the last decade, large-scale, organized (generally dipolar) magnetic fields with a strength between 0.1 and 20 kG were detected in dozens of OB stars. This contribution reviews the impact of such magnetic fields on the stellar winds of O-stars, with emphasis on variability and X-ray emission.}

\baselineskip 12pt

\section{Introduction}
It has long been suspected that massive stars should possess magnetic fields. Indeed, pulsars, or even more extreme magnetars, are not the remnants of low-mass stars! However, the detection of magnetic fields is difficult in massive stars: spectral lines are few in number and quite broad, hiding Zeeman splitting; furthermore, emission arising in the wind or contamination by bright companions may dilute the signal. For a long time, only indirect observational evidence could thus be put forward: synchrotron radio emission, peculiar phenomena such as line profile variability (in particular discrete absorption components, DACs), or very hot X-ray emission. Their presence was however debated, as massive stars lack the convective envelopes responsible for the magnetic dynamo in late-type stars.

In the last decade, a revolution took place: magnetic fields were detected in O-stars. This was the outcome of sensitive spectropolarimetric surveys. In such studies, the Zeeman splitting is not detected per se, but the associated circular polarisation across line profiles is measured. To this aim, the normalized Stokes $V/I$ profiles are calculated from the set of spectra obtained for different angles of the retarder wave plate; for sanity checks, best practices include the calculation of a diagnostic ``null'' profile \citep[see ][for details]{don97,bag09}. 

High-resolution spectrographs show in detail the change in $V/I$ across the line profile, leaving little doubt on the detection. When needed, the signal-to-noise ratio can be increased by combining the common $V/I$ signal from many individual spectral lines \citep[LSD technique,][]{don97}. The variation of this $V/I$ profile with the rotation period then allows to derive the magnetic field configuration, either from measurements of the mean longitudinal magnetic field (e.g. \citealt{don06tau}), or by forward or inverse modelling of the profile variations (e.g. Magnetic Doppler Imaging, \citealt{koc02} and references therein).

High-resolution cannot always be used, especially when the stars are faint. Low-resolution instruments do not allow the determination of a detailed $V/I$ profile, but the longitudinal magnetic field (\bz) can still be estimated \citep{bag02}. Recently, doubts were expressed on the reality of some magnetic field detections made with FORS (e.g. unconfirmed claims, inconsistencies between measurements). \citet{bag12} thus undertook a homogeneous reduction of all spectropolarimetric FORS1 data, showing the implications of data reduction choices. The conclusion is that FORS is a reliable instrument for magnetic field searches, but that a detection level of 5--6$\sigma$ is required to avoid false alarms. 

\section{The prototype: $\theta^1$\,Ori C}
$\theta^1$\,Ori C (O7V) is the brightest and hottest star in the Orion nebula's trapezium. It is a visual binary, and interferometric data yield an orbit with P=11yr, e=0.5-0.6, and a mass ratio of about 0.2 \citep{kra09}. \citet{leh10} suggested $\theta^1$\,Ori C to be a triple system, maybe with a 1:4 resonance with the rotation period, but that needs to be confirmed.

Changes in the spectrum of $\theta^1$\,Ori C were first reported by \citet{con72}, and then found to be periodic ($P$=15.4d, \citealt{sta96}). Photospheric lines vary with a smaller amplitude than emission lines, and with maximum absorption appearing with maximum emission; UV lines such as CIV$\lambda\lambda$1548,1550 show increased absorption on the blue wing (and slightly decreased absorption on the red wing) when emission lines are weakest. Those variations were interpreted in the framework of the magnetic oblique rotator model: a dipolar magnetic field channels the stellar winds from the two opposite hemispheres towards the magnetic equator, forming a disk-like feature which is alternatively seen edge-on and face-on as the magnetic and rotational axes are different. Maximum H$\alpha$ and HeII$\lambda$4686 emissions correspond to the equatorial feature seen face-on. The presence of a magnetic field was finally confirmed by spectropolarimetric measurements of \citet{don02}: $B_d=1.1$\,kG, $\beta=42^{\circ}$ for $i=45^{\circ}$. It was the first detection of a magnetic field in an O-star. Recent data have confirmed these properties \citep{wad06,hub08}. 

The wind flows from both hemispheres, channeled by the magnetic field, collide at the equator, heating the gas to high temperatures. This generates hard X-ray emission \citep{bab97,udd02}. Already in the Einstein and ROSAT era's, $\theta^1$\,Ori C was known for its peculiar X-ray emission but recent Chandra data revealed more details \citep{sch00,gag05}: the X-ray emission appears thermal and dominated by $\sim3$\,keV plasma, it is also bright (\loglxlb$\sim-6.0$) and modulated in phase with the 15d period; the hot plasma is close to the star and moving slowly (narrow X-ray lines, X-ray formation region at about 2$R_*$). All these properties agree well with the expectations from the oblique rotator model, so that the star often plays the role of a prototype.  However, while detailed 2D and 3D simulations reproduce well the H$\alpha$ variations \citep{udd13}, a few observations cannot be readily reproduced (double-peaked minimum in H$\alpha$ variations, X-ray velocity shifts and absorption, behaviour of UV lines - see e.g. \citealt{udd08}).

\section{The magnetic group: Of?p stars}
The Of?p category was defined by \citet{wal72} for stars presenting peculiarities, especially strong emission of the CIII$\lambda$4650 lines. Three stars in our Galaxy were then members of this category: HD108, HD148937, and HD191612. It now appears that strong CIII emission alone is not sufficient to define an Of?p star (the stars showing only this feature are now called Ofc stars, Walborn et al. 2010). The Balmer hydrogen lines of Of?p stars have a composite nature, with narrow emissions superimposed on the broader stellar components; HeI lines have asymmetric or P\,Cygni profiles; UV lines, such as SiIV near 1400\AA, also appear peculiar, unlike what is expected for Of supergiants; sometimes, there is also emission in the SiIII triplet around 4568\AA\ \citep{naz08rvmx}.

The most important characteristics of these objects is, however, their periodic variability.  After the first investigations for HD108 \citep{vre79}, detailed analyses awaited the 21st century. Dramatic changes were then found in the Balmer and HeI lines of HD108 \citep{naz01} and HD191612 \citep{wal03}. Dedicated spectral monitorings revealed a period for the variability: about 55yrs for HD108 \citep{naz01,naz06} and 538d for HD191612 \citep{wal04,how07}. Photometric variations with the same period were also found \citep{bar07,koe02,wal04}, with a maximum luminosity corresponding to the maximum emission state. The variability of the last "historical" case, HD148937, appears similar in nature to those of HD108 and HD191612 though with a much smaller amplitude and a much shorter period \citep[only 7d,][]{naz08,naz10}. 

In 2006, the detection of a magnetic field was reported for HD191612 \citep{don06}. Further monitoring showed the field to be dipolar, with a strength $B_d\sim 2.5$\,kG and inclination $\beta=67^{\circ}$ for $i=30^{\circ}$ \citep{wad11}, in agreement with the results of a toy model reproducing the H$\alpha$ variations \citep{how07}. Longitudinal magnetic fields of 100--300\,G (or $B_d\sim 1$\,kG) were also detected for both HD108 \citep{mar10} and HD148937 \citep{hub08,hub11}. The low-amplitude variability of HD148937 was linked to the particular geometry of the system \citep{naz10,wad12hd} while the long period of HD108 is suggested to be a consequence of magnetic braking \citep{mar10}. 

Since the "historic" detections, a few additional Of?p stars were identified both in the Magellanic Clouds and in the Galaxy \citep{hey92,wal00,mas01,wal10}. Magellanic clouds' objects cannot yet be studied in detail with spectropolarimeters, but magnetic fields were searched, and found, for the two new Galactic cases: CPD$-$28$^{\circ}$2561 \citep[$P\sim70$d, $B_d\sim 1.5$\,kG,][ Barba et al. in prep]{hub11,hub12,hub13}; NGC1624-2 \citep[$P\sim158$d, $B_d\sim 20$\,kG, the strongest magnetic field on record - Zeeman splitting is even detected for CIV$\lambda\lambda$5801,5814 lines,][]{wad12ngc}.   

Turning to high energies, the three "historical" Of?p cases display very similar spectra with strong overluminosities (\loglxlb$\sim-6.1$, \citealt{naz04,naz07,naz08}). Hints of overluminosities in the X-ray range have been detected for CPD$-$28$^{\circ}$2561 (\citealt{hub13}, Naz\'e et al., in prep) and NGC1624-2 (\loglxlb$\sim-6.4$, \citealt{wad12ngc}). Furthermore, the X-ray emission of HD191612 appears modulated with the 538d period \citep{naz07,naz10} and the narrow lines expected for a confined wind were finally detected in HD148937 for the high-Z elements using Chandra \citep{naz12hd}. These properties make the Of?p stars similar to $\theta^1$\,Ori C except for one ingredient: the general softness of the X-ray emission, also seen in some magnetic B-stars (e.g. \citealt{osk11}), but in conflict with model predictions \citep{naz10}. 

Ultraviolet data also yielded surprising results. HD108 has been observed with IUE close to the maximum emission phase, and with HST-STIS close to minimum emission phase. These UV spectra show only moderate variations, larger than observed in non-magnetic single O-stars, but much smaller than the drastic changes detected in the optical domain \citep{marco12}. This can probably be explained by the larger formation zones of the UV lines (well above the Alfven radius) compared to those of the Balmer H lines . Strangely, the UV lines display a smaller absorption when the dense equatorial region is seen edge-on, contrary to naive expectations (disk seen edge-on implying more absorption) and to the case of $\theta^1$\,Ori C (see above). A HST-STIS monitoring of HD191612 \citep{marco13} revealed intriguing features similar to HD108, with only two differences: UV line profiles are never saturated, and SiIV lines displays a behaviour opposite to that of CIV and NV lines. This dichotomy could be qualitatively reproduced by MHD simulations, and appears to be due to the strength of lines (strong vs weak) considered. 

\section{Other objects}

\subsection{HD57682}
This O9IV star was found to be magnetic by \citet{gru09}, and a more detailed study was published afterwards \citep{gru12}: $P$=63.6d, $\beta=79-88^{\circ}$, and, for the favored inclination $i>30^{\circ}$, $B_d<1.5$\,kG. The particularity of this object is its geometry: both magnetic poles are alternatively seen, and the equatorial "disk" is thus seen face-on twice per period, leading to double-peaked variations in the strengths of some disk-related lines (e.g. H$\alpha$). Simple `toy' models and MHD simulations are able to reproduce these variations in line intensity, but fail to reproduce the associated radial velocity changes. The latter could be due to asymetries in the "disk" or to an offset between the dipole and the star's center. 

Other lines in HD57682 also vary, but in a different way (single-peaked, not double-peaked). Several explanations were advanced (binarity, pulsations, chemical spots) but discarded. The only remaining possibility is that the observed changes are also linked to the magnetic field: indeed, there is a smooth transition amongst Balmer lines, with H$\alpha$ showing strong double-peaked EW variations,while H$\gamma$ changes are single-peaked and H$\beta$ variations are in between. A full 3D modelling of the system would certainly help better understand this peculiar behaviour.

I have obtained X-ray data of HD57682 (Naz\'e et al., in prep). Its X-ray spectrum is slightly harder than usual, and the X-ray emission is slightly overluminous (\loglxlb$\sim -6.4$). It thus appears as a less extreme case than Of?p stars, $\theta^1$\,Ori C, or Tr16-22 (see below).

\subsection{The first massive magnetic (close) binary}
HD47129, also known as Plaskett's star, is a massive binary composed of an O8III/I primary and an O7.5V/III secondary in a 14d circular orbit \citep{lin08}. The system shows several peculiarities: the secondary is rapidly rotating whilst the primary has a much lower rotational velocity; the primary is brighter than the secondary despite having a mass similar to its companion; and the abundances of both components are anomalous (primary strongly N-enriched and C-depleted, secondary N-depleted and He-enriched). These peculiar properties led to the conclusion that Plaskett's star is a post Roche lobe overflow system \citep{lin08}. 

Recently, a magnetic field was detected for the secondary component \citep{gru13}. The measured longitudinal fields vary between $+680$ and $-810$G, with errors $<$200G: if the field is dipolar, its strength then amounts to $\sim$2\,kG. The detections of a bright, hard and variable X-ray emission and of a flattened wind around the secondary component by \citet{lin08} are compatible with magnetically confined winds. Further monitoring (Grunhut et al. in prep) led to the detection of a large inclination (as in HD57682) and possibly of the rotational period, in line with the frequency 0.823\,d$^{-1}$ detected by Corot \citep{mah11}. Because it is a binary, Plaskett's star appears as a unique laboratory for testing several phenomena. Indeed, the past Roche lobe overflow has modified the stellar structures and dynamics. It is thus expected that the magnetic field configuration has been affected by the event, but the details are still unknown. Furthermore, the stars are close to each other, with the primary at least partially inside the secondary's magnetosphere. Also, the confined wind region around the secondary is so large that it encompasses the expected apex of the (potential) wind-wind collision. The question then arises of the interplay between the secondary's magnetic field and the primary's wind (deflected or channeled ?). In this framework, the lack of strong emission at visible wavelengths associated with the secondary may appear puzzling. Additional investigations, both observational and theoretical, should thus be undertaken to understand this magnetic "Rosetta stone".

\subsection{The first X-ray identified magnetic O-star}
Most O-stars display an intrinsic soft X-ray emission following $L_{\rm X}\sim 10^{-7}\times L_{\rm BOL}$ (\citealt{naz11} and references therein). In the past, overluminosities were often thought to be linked to wind-wind collisions in binaries, but many massive binaries now appear to display a "normal" \lxlb\ ratio (\citealt{naz11} and references therein) and magnetic confinement was proposed as a second source of hard X-rays  \citep{bab97}. Distinguishing between the two mechanisms requires a monitoring, to see whether changes are phased with the orbital period (colliding winds) or the rotation period (magnetic confinement) - see e.g. the case of HD\,191612  \citep{naz10}.

During a Chandra survey of the Carina nebula, several objects were found to display an overluminosity and/or hard X-ray emission, triggering a spectropolarimetric campaign. One of the most promising targets was Tr16-22, whose X-ray emission is hard , bright, and variable. Moreover, its late-type (O8.5V) renders an X-ray bright wind-wind collision unlikely. A longitudinal field of $\sim-500$\,G (6$\sigma$ detection) was detected, together with narrow lines associated with slow rotation, as in Of?p stars \citep{naz12car}. Today, the monitoring of this object continues in order to pinpoint the magnetic configuration of Tr16-22. The example of Tr16-22 shows that selecting targets using X-rays is an efficient tool, to be used along with other indirect indicators (e.g. UV or visible peculiarities).

\section{Relation with other magnetic stars}
It is interesting to compare the magnetic O-stars to their B-star colleagues  \citep{pet13}. When considering the magnetospheres of massive stars, there are a few key parameters. The first one is the magnetic confinement, defined as $\eta_*=B_{eq}^2 R_*^2/\dot M v_{\infty}$  \citep{udd02}. It compares the magnetic to wind kinetic energy density: values much larger than unity indicate high degrees of confinement. This parameter is linked to the Alfven radius $R_A$, which defines the size of the magnetosphere: magnetic loops smaller than $R_A$ remain closed and material in this region is forced to corotate with the star. The second important parameter is the stellar rotation rate: with negligible rotation (i.e. $R_A<$ the Keplerian corotation radius $R_K$), the centrifugal support is weak, and material in the magnetosphere cannot resist gravity and falls back onto the star, though some outflow also exists at larger radii. In this case, the star has a complex {\it dynamical magnetosphere}, this dynamic behaviour explaining the short-time or cycle-to-cycle variability observed for Of?p stars. At higher rotation speeds (i.e. $R_A>R_K$), the centrifugal support is higher, and the material trapped in the magnetosphere can then accumulate between $R_A$ and $R_K$, forming a dense {\it centrifugal magnetosphere}. In practice, because of their high mass-loss rates enabling magnetic braking, magnetic O-stars (except for Plaskett's star) display low rotation, hence dynamical magnetospheres. Consequently, magnetic O-stars show H$\alpha$ emission and X-ray overluminosities, whereas such features are only seen in the most extreme B-stars (strong magnetic field, fast rotation) because enough material needs to be accumulated from the low mass-loss rates of B-stars before an emission can be detected. 

In this context, it may also be worth taking a look at the abundances. Theoretically, the impact of magnetic fields on the rotational mixing in stellar interiors is unclear \citep{mey11}. Observationally, there is a higher incidence of N-excess in magnetic B-stars, but there is no one-to-one relation (detection in some cases but not in others, \citealt{mor08,mor11,mor12,prz11}). The situation is similar in O-stars: N-overabundance in the three "historical" Of?p stars \citep{naz08rvmx,mar12} but not in NGC1624-2 and CPD$-$28$^{\circ}$2561 (\citealt{wad12ngc}, Barba et al. in prep); inconclusive results found for HD57682 \citep{kil92,mor11,mar12}. A larger sample of clearly magnetic (and clearly non-magnetic) OB-stars is now needed to draw firm conclusions about mixing processes.

\section{Summary and Conclusions}
After having long been suspected to exist, magnetic fields have finally been detected in O-stars over the last decade. However, only a handful of stars (5--15\%) were found to harbour a strong, dipole-like magnetic field, a fraction similar to AB stars  \citep{hub11,wad12aspc}. Of?p stars clearly form a class of magnetic O-stars; \citet{hub13} suggested that runaway stars may constitute another one. 

In the rare cases of strongly magnetic O-stars, variability seems to be the rule, whatever the wavelength (X-ray, UV, or visible domain). The changes are mostly related to the varying viewing angle on the magnetically-confined wind material, as the magnetic axis is inclined on the rotational axis. Depending on the exact geometry, the variations of the line intensities may be small (HD148937) or large (HD191612), single-peaked (HD191612) or double-peaked (HD57682). While the optical and UV spectra are now quite well understood, the softness of the X-ray emission of many magnetic O-stars remains a puzzle, as for B-stars (e.g.  \citealt{ign10}). 

The detection of a weak field in $\zeta$\,Ori  \citep{bou08} has opened the door to another category of magnetic O-stars, even if this particular detection is now questioned (Neiner et al. in prep.). Strong dipolar fields may be rare, but weak, non-dipolar fields could be widespread \citep[current observations excluding only strong small-scale fields, ][]{koc13}. Such complex fields have been proposed to explain the recurrent DACs in O-star lines (e.g. $\lambda$\,Cep) and the possible presence of hard X-ray emission close to the photosphere \citep{waldron09}. 

Another domain of interest is the presence of magnetic fields in massive binaries, which should be explored in the near future: the fields of the binary components could interact, potentially with some impact on the wind-wind collision, and they can be modified by binary interactions (e.g. mass transfer). 

Future instrumentation should thus not only confirm the current low-significance detections and detect more cases of strongly magnetic objects, including binaries, but also seek to discover stars with very weak fields and/or complex magnetic topologies. In parallel, theoretical developments are also needed to better understand these massive magnetospheres, especially their high-energy emission. The story of magnetic field studies in the massive stars' community has only just begun.

\bigskip
{\it Acknowledgements.} I thank the SOC and LOC (incl. Igor) for their invitation and help. I acknowledge support from ULg and FNRS, and careful reading by G. Rauw, T. Morel, A. ud-Doula, and G. Wade.

\end{document}